\documentclass[amsmath,amssymb,aps,prx,reprint,superscriptaddress]{revtex4-2}
\usepackage{amsmath}
\usepackage[table]{xcolor}
\usepackage{amsfonts,amssymb}
\usepackage{graphicx}
\usepackage{hyperref}
\usepackage{academicons}
\usepackage{multirow}
\usepackage[version=4]{mhchem}
\usepackage{makecell}
\usepackage{pgf}
\usepackage{booktabs}
\usepackage[english]{babel}

\newcommand{\CellMin}{0}
\newcommand{\CellMax}{0.4}
\colorlet{cmin}{green!10!white}  
\colorlet{cmed}{yellow!20!white} 
\colorlet{cmax}{red!20!white}    
\newcommand{\ColorCell}[1]{%
  \begingroup
  \pgfmathsetmacro{\val}{#1}%
  \pgfmathsetmacro{\norm}{(\val-\CellMin)/(\CellMax-\CellMin)}%
  \pgfmathsetmacro{\norm}{max(min(\norm,1),0)}%
  \pgfmathsetmacro{\pct}{\norm*100}%
  \pgfmathtruncatemacro{\pctint}{round(\pct)}%
  \ifnum\pctint<50
    \pgfmathtruncatemacro{\mix}{round(100 - 2*\pctint)}%
    \edef\temp{\noexpand\cellcolor{cmin!\mix!cmed}}%
  \else
    \pgfmathtruncatemacro{\mix}{round(200 - 2*\pctint)}%
    \edef\temp{\noexpand\cellcolor{cmed!\mix!cmax}}%
  \fi
  \temp #1%
  \endgroup
}

\begin{document}

\title{Interfacial melting as a thermodynamic indicator of solid-state synthesizability}

\author{Zihan Zhang}
\thanks{These authors contributed equally.}
\affiliation{Department of Materials Science and Engineering, National University of Singapore, Singapore}

\author{Mengyi Chen}
\thanks{These authors contributed equally.}
\affiliation{Department of Mathematics, National University of Singapore, Singapore}

\author{Qianxiao Li}
\affiliation{Department of Mathematics, National University of Singapore, Singapore}
\affiliation{Institute for Functional Intelligent Materials, National University of Singapore, Singapore}

\author{Peichen Zhong}
\email[]{zhongpc@nus.edu.sg}
\affiliation{Department of Materials Science and Engineering, National University of Singapore, Singapore}
\affiliation{Institute for Functional Intelligent Materials, National University of Singapore, Singapore}

\date{\today}

\begin{abstract}
Computational materials discovery commonly ranks candidate materials by their thermodynamic stability on the formation energy convex hull, yet many predicted-stable phases resist synthesis. We propose that solid-state synthesizability through interfacial-melt-mediated routes requires an additional thermodynamic condition: the interfacial melt at the target composition must itself remain locally stable against spinodal decomposition.
We examine this in the classical Fe--B system, where thermodynamically stable FeB$_4$ has been reported under high-pressure synthesis but not in low-pressure synthesis attempts.
Using melt--quench molecular dynamics driven by a fine-tuned machine-learning interatomic potential, we find that, at ambient pressure, the B-rich interfacial melt near the FeB$_4$ composition develops a concave free-energy landscape, signaling a demixing instability that is corroborated by the concentration--concentration structure factor and correlated with low-energy icosahedral and pentagonal-pyramidal boron motifs. In contrast to FeB$_4$, metastable Fe$_3$B and Fe$_{23}$B$_6$ remain synthesizable because their corresponding melts are stable.
Applied pressure introduces a convex $PV$ contribution that strongly suppresses this instability, reducing the curvature at the FeB$_4$ composition to within the uncertainty of our fit at 1800~K, consistent with the experimental synthesis boundary. Comparison with CrB$_4$ further shows that weaker melt instability correlates with easier experimental synthesis. Interfacial-melt stability, which atomistic simulations can assess via the low-$k$ concentration--concentration structure factor, is thus proposed as a practical thermodynamic screening descriptor of synthesizability for AI-assisted materials discovery.
\end{abstract}


\maketitle

\section{Introduction}
Computational materials discovery commonly ranks candidate crystals by thermodynamic stability, yet a phase on the formation energy convex hull is not necessarily synthesizable \cite{Riebesell_2025, Yuan_2026, Lejaeghere_2016, horton_accelerated_2025, Oganov_2019}. AI-accelerated workflows that combine crystal structure search \cite{pickard_2011, oganov_2006, wang_2010} and generative models \cite{Zeni_2025, Luo_2025, zhong_2025} with foundation interatomic potentials trained on extensive quantum-mechanical data have vastly expanded the catalog of predicted stable inorganic materials \cite{wood_uma_2025, batatia_foundation_2025}, in which synthesizability is implicitly equated with lying on the formation energy convex hull. In practice, although thermodynamic stability correlates well with synthesizability \cite{miura_selective_2020, szymanski_2023}, many predicted-stable phases have yet to be synthesized. The missing physics is the synthesis process itself: hull stability, whether evaluated from formation energies or finite-temperature free energies, is an equilibrium property of the final crystal and says nothing about the interfacial melting and nucleation through which that crystal must form.

FeB$_4$ offers a sharp illustration of this gap. 
Density functional theory calculations at the PBE level predict FeB$_4$ to be a thermodynamically stable, phonon-mediated superconductor \cite{Kolmogorov_2010}, and with vibrational free energy included FeB$_4$ lies on the ambient-pressure hull above about 1000~K \cite{10.1063/1.3556564}. The same conclusion holds at the r$^2$SCAN level (see SI).
Yet bulk FeB$_4$ has not been identified in that ambient-pressure phase diagram \cite{Zhang_2019}, and has been reported under high-pressure synthesis but not in low-pressure attempts \cite{Gou_2013, Wang_2015_FeB}. The discrepancy is telling: at synthesis temperatures FeB$_4$ lies on or within about 10 meV/atom of the ambient-pressure hull at both PBE and r$^2$SCAN levels of theory (see SI), so the thermodynamic driving force of the product alone cannot explain the pressure dependence of its synthesis. The missing piece is the pathway by which it forms \cite{Chen_2024}.

\begin{figure*}[htbp]
    \centering
    \includegraphics[width=\textwidth]{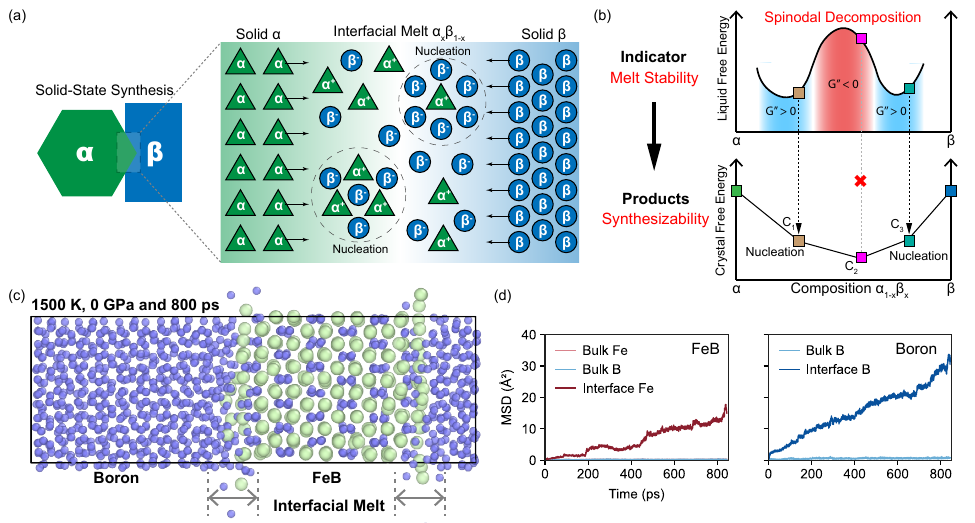}
    \caption{
    (a) Schematic of solid-state synthesis from precursors $\alpha$ and $\beta$. The interfacial melt $\alpha_{1-x}\beta_x$ that forms at the interface provides the nucleation environment for the target product.
    (b) Thermodynamic indicator for synthesizability. The sign of the free-energy curvature partitions the $\alpha_{1-x}\beta_x$ melt into locally stable ($G''>0$) and spinodally unstable ($G''<0$) regions; because unstable compositions decompose spontaneously (C$_2$), viable melt compositions are restricted to the stable or metastable regions (C$_1$, C$_3$). Nucleation of the product into its crystalline form proceeds only from a locally stable or metastable melt (C$_1$, C$_3$); a target composition inside the spinodally unstable region (C$_2$) is inaccessible through this direct melt-mediated route.
    (c) Final configuration at 1500~K and 0~GPa from two-phase MD simulation of interfacial melting between FeB and $\alpha$-boron. Green and blue spheres represent Fe and B atoms, respectively. For visual clarity, periodic boundary conditions have been removed in the displayed snapshots so that atoms outside the original lattice boundaries correspond to those that have diffused across the interface.
    (d) Mean square displacement (MSD) of atoms in the interfacial and bulk regions of FeB (left) and $\alpha$-boron (right), obtained from two-phase molecular dynamics simulations. The markedly larger MSD at the interface indicates enhanced atomic mobility and incipient interfacial melting.
    }
    \label{fig:figure1}
\end{figure*}

Solid-state synthesis is widely understood to proceed through reactive interfaces, where interfacial melting provides a general pathway for the reaction  \cite{Miura_2021, mcdermott_assessing_2023, gallant_cellular_2025}. Solid interfaces typically have a higher energy density than the bulk, and the melting point at the interface between solid reactants is lower than in the bulk. Consequently, absent special kinetic behaviors such as temperature-induced phase transitions or superionic states in the reactants, interfacial melting constitutes a general mechanism for solid-state reactions. When two solid precursors are heated in contact, they react across their shared interface, where in many systems a thin transient liquid forms by eutectic melting \cite{hu_resolving_2023, Karan_2026} (Fig.~\ref{fig:figure1}(a)). The target crystal nucleates and grows from this interfacial melt, whose composition varies continuously across the reaction zone. The melt, not only the final crystal, therefore constrains which phases can form. We propose that synthesizability via this interfacial-melt-mediated route carries a thermodynamic indicator: the interfacial melt at the target composition must itself be locally stable. A melt is locally stable only where its free energy is convex in composition space. When the free-energy curve is concave, the melt can lower its energy by spontaneously separating into two compositions through spinodal decomposition \cite{Unger_1984, Park_2024}. A target composition lying in such a spinodally unstable window cannot persist as the homogeneous parent liquid required for nucleation, so the corresponding crystal is inaccessible by this route despite lying on the convex hull. For Fe--B, this offers a thermodynamic rationale for the FeB$_4$ puzzle: at low pressure the B-rich melt is spinodally unstable near the FeB$_4$ composition, whereas applied pressure adds a convex $PV$ contribution that strongly suppresses the instability, matching the observed synthesis boundary.

\section{Results and Discussion}

\begin{figure*}[htbp]
    \centering
    \includegraphics[width=\textwidth]{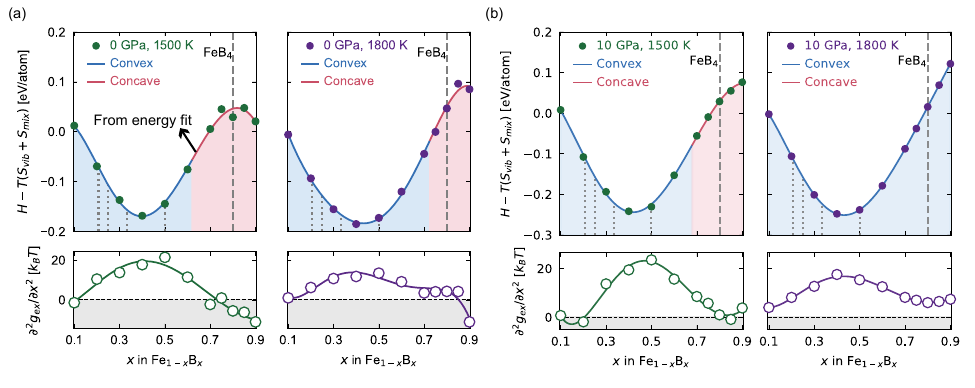}
    \caption{
    Two estimators of melt stability for the Fe$_{1-x}$B$_{x}$ melt versus composition $x$ at 1500 and 1800~K, for (a) 0~GPa and (b) 10~GPa. Top: free energy $H-T(S_\mathrm{vib}+S_\mathrm{mix})$, with regions colored by the sign of its curvature from a polynomial fit, classifying each composition as convex or concave (blue and red regions in Fig.~\ref{fig:figure1}(b)).
    Bottom: the structure-factor-derived excess curvature $\tfrac{1}{k_{\mathrm B}T}\,\partial^2 g_{\mathrm{ex}}/\partial x^2$ from Ornstein–Zernike extrapolation; a negative value indicates a demixing tendency (gray).
    FeB$_4$ is marked by a dashed line and the ambient-pressure phases by dotted lines (Fe$_{23}$B$_6$, Fe$_3$B, Fe$_2$B, FeB) \cite{quirinale_2015, khan_1982, Zhang_2019}.
    }
    \label{fig:figure2}
\end{figure*}

\noindent\textbf{Two-phase MD of solid-solid interfaces.}
The criterion above presupposes that a liquid-like layer forms where the reactants meet. To test this for Fe--B, we constructed a two-phase supercell by joining the (010) surface of FeB with the (100) surface of $\alpha$-boron, the reactant pair of the FeB $+$ 3B synthesis route, and ran NPT molecular dynamics (MD) at 1500~K and 0~GPa with the fine-tuned potential, a temperature well below the melting points of FeB and boron \cite{okamoto_B_Fe_2004}. Fig.~\ref{fig:figure1}(c) shows the configuration after 800~ps. Both slabs retain their lattice order in the interior, whereas a disordered zone a few atomic layers thick has developed at each interface. Fig.~\ref{fig:figure1}(d) quantifies this with the mean square displacement (MSD) of atoms initially located at the interface and in the bulk of each crystal. The interfacial Fe and B atoms show MSDs that grow steadily with time, a signature of liquid-like diffusion, whereas the bulk atoms in both crystals remain near their lattice sites. The interface therefore melts preferentially, as expected from the eutectic effect, and the resulting layer is a Fe--B liquid-like structure whose composition varies from FeB-rich to B-rich across the reaction zone.

This observation also suggests how the interfacial melt should be modeled. The criterion requires the free energy as a function of composition, $G(x)$, a local property of each compositional slice of the interfacial layer. 
Two-phase MD cannot supply it: the cell is phase-separated by construction, so it cannot show whether a homogeneous liquid-like structure of a given composition would demix. The time for the interfacial liquid to homogenize and for products to precipitate from it lies far beyond MD time scales, e.g., on the order of minutes in in situ diffraction studies of solid-state reactions \cite{Szymanski_2024, Karan_2026}.
We therefore separate the interfacial structure conceptually into its compositions $x$ and simulate each as a homogeneous Fe$_{1-x}$B$_x$ melt at fixed temperature and pressure with a melt--quench protocol, which erases the memory of the initial structure and yields an equilibrated liquid at each state point. 
The curvature of $G(x)$ and the concentration fluctuations of these melts then give the local stability of the interfacial liquid at every composition it passes through. 
This treats the interfacial layer as a bulk liquid-like structure and neglects confinement and gradient effects.

\noindent\textbf{Free energy landscape across composition.}
This criterion can be made precise. Confined to the interface, the melt free energy is a continuous function of composition, $G(x)$ for $\alpha_{1-x}\beta_x$, whose curvature partitions the composition axis (Fig.~\ref{fig:figure1}(b)). A composition is spinodally unstable where
\begin{equation}
    G'' = \frac{d^2G}{dx^2} < 0 \quad (\text{spinodal}),
\end{equation}
so that infinitesimal concentration fluctuations grow without a nucleation barrier and the melt separates into $\alpha$-rich and $\beta$-rich domains; compositions with $G'' > 0$ are locally stable or metastable and demix only by nucleation. 
Because direct nucleation of the product presupposes a homogeneous parent melt at its stoichiometry, a target inside the spinodal window is inaccessible through this direct route unless another pathway bypasses the melt. The criterion is therefore necessary but not sufficient for synthesis along this route.
For multicomponent melts $\{x_i\}$, the criterion generalizes to positive-definiteness of the free-energy Hessian $G_{ij} = \partial^2 G / \partial x_i \partial x_j$.

We test this criterion for the Fe--B interfacial melt by evaluating the curvature $G''$ in two complementary ways (see methods in the Supporting Information, SI). 
The first extracts the free-energy landscape from the melt energetics. We performed melt--quench MD to generate the Fe$_{1-x}$B$_x$ melts, and sampled the trajectories from MD simulations across the composition range ($x = 0.1$--$0.9$) in the NPT ensemble at 0 and 10~GPa, driven by a Fe--B fine-tuned MACE-MH-1 potential with the r$^2$SCAN head \cite{barros2026open, batatia_cross_2025, kaplan_foundational_2025}. 
For each trajectory we evaluate the free energy $H-T(S_\text{vib} + S_\text{mix})$, referenced to the elements via $E_{\mathrm{norm}}(\mathrm{Fe}_{1-x}\mathrm{B}_x) = E(\mathrm{Fe}_{1-x}\mathrm{B}_x) - (1-x)E(\mathrm{Fe}) - xE(\mathrm{B})$. Here $H$ is the trajectory-averaged enthalpy, $S_\mathrm{vib}$ follows from the vibrational density of states \cite{lin_two-phase_2003}, and the configurational entropy is approximated by its ideal-solution value $S_\mathrm{mix} = -k_{\mathrm B}(x_\text{Fe} \ln x_\text{Fe} + x_\text{B} \ln x_\text{B})$. Because the ideal value is an upper bound on the true configurational entropy of a clustering melt, this approximation over-stabilizes the homogeneous state, so any residual concavity is a conservative indicator of instability. Differentiating a polynomial fit yields $G''(x)$, classifying each composition as convex or concave (blue and red regions in Fig.~\ref{fig:figure2}(a,b)). 

At 0~GPa and 1500~K, the landscape is convex on the Fe-rich side but shows pronounced concavity near the FeB$_4$ composition (Fig.~\ref{fig:figure2}(a)), indicating that the FeB$_4$ ($x=0.8$) melt is locally unstable and can undergo spinodal decomposition, inhibiting crystallization of the thermodynamically stable FeB$_4$ \cite{Kolmogorov_2010}. The four iron borides that form at ambient pressure (the stable Fe$_2$B and FeB \cite{Zhang_2019}, and the metastable Fe$_{23}$B$_6$ \cite{quirinale_2015} and Fe$_3$B \cite{khan_1982}) all lie within the locally stable region. Raising the temperature narrows the unstable window, leaving the FeB$_4$ melt only marginally unstable at 1800~K. Pressure changes this picture qualitatively: at 10~GPa the concavity is markedly reduced at 1500~K and falls within the uncertainty of the fit at 1800~K (Fig.~\ref{fig:figure2}(b)).

Interestingly, the metastable Fe$_3$B and Fe$_{23}$B$_6$ phases provide a contrasting case within the same Fe--B system \cite{Kolmogorov_2010}. 
Both phases lie above the convex hull, whereas FeB$_4$ lies on or close to it. Nevertheless, they can be reliably synthesized because their B-deficient melts are stable, highlighting that crystalline thermodynamic stability alone does not determine solid-state synthesizability.

\begin{figure}[htbp]
    \centering
    \includegraphics[width=\columnwidth]{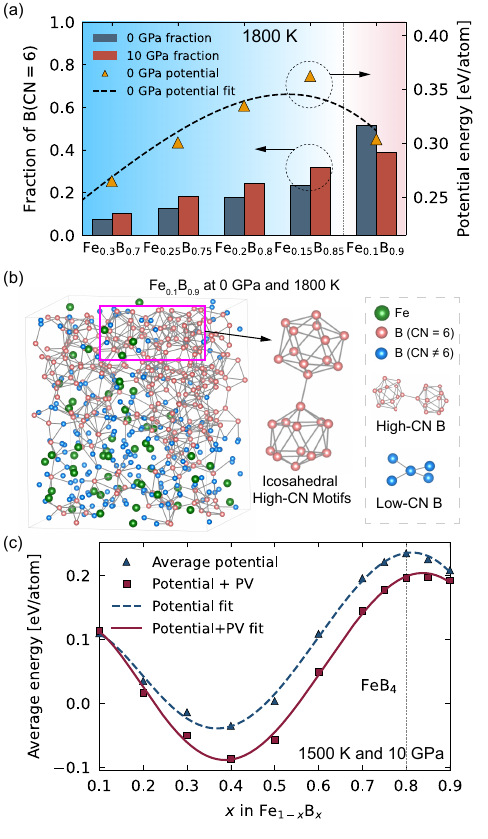}
    \caption{
    Structural and energetic properties of B-rich Fe$_{1-x}$B$_{x}$ melts.
    (a) Fraction of six-coordinated (CN=6, B--B cutoff of 2.37 \AA\ from the first minimum of the B–B radial distribution functions, see Figs. S18--S28) B atoms at 0~GPa (dark blue) and 10~GPa (dark red), left axis, and average potential energy at 0~GPa (yellow triangles and fit), right axis, at 1800~K.
    (b) MD snapshot of Fe$_{0.1}$B$_{0.9}$ at 0~GPa and 1800~K; green, orange, and blue spheres are Fe, B(CN=6), and B(CN$\neq$6), with only CN=6 B--B bonds shown.
    (c) Average potential energy with (red squares, solid) and without (blue triangles, dashed) the $PV$ correction at 1500~K and 10~GPa.
    }
    \label{fig:figure3}
\end{figure}

\noindent\textbf{Structure factor from MD simulations.}
The energy-based estimate is useful but not sufficient: it infers curvature by differentiating an empirical polynomial fit to finite-time energetics from non-equilibrium MD simulations. Near a spinodal, the trajectories can already contain incipient concentration fluctuations or local demixing. We therefore also assess stability from the concentration fluctuations of the liquid itself, which provide a structure-based diagnostic and do not rely on differentiating the fitted energy surface.
Within classical density-functional theory, the free energy splits as $g = g_{\text{id}} + g_{\text{ex}}$ into ideal and excess parts, and the thermodynamic factor is
\begin{equation}
\begin{aligned}
    \Gamma & = 1 + \frac{x_\alpha\,x_\beta}{k_{\mathrm B}T}\,\frac{\partial^2 g_{\mathrm{ex}}}{\partial x^2} = \frac{x_\alpha x_\beta}{S_{\text{cc}}(\mathbf{k}\rightarrow 0)} \\
    & = \frac{1}{x_\beta S^0_{\alpha\alpha} + x_\alpha S^0_{\beta\beta} - 2\sqrt{x_\alpha x_\beta }S^0_{\alpha\beta} }
\end{aligned}
\end{equation}
where $S^0_{\alpha\beta}$ is the zero-wavevector partial structure factor and $x_{\alpha/\beta}$ is the molar fraction of component $\alpha/\beta$ \cite{barrat_basic_2003}. Here $\mathbf{k}$ denotes the discrete wavevectors of the periodic cell, $k=|\mathbf{k}|$ its magnitude, and $L_{\max}$ the longest cell edge, which sets the smallest resolved wavenumber $k_{\min}=2\pi/L_{\max}$. The spinodal corresponds to $\Gamma \to 0$, where the concentration--concentration structure factor $S_{\text{cc}}(\mathbf{k}\rightarrow 0) = x_{\alpha} x_{\beta}/\Gamma$ diverges.
Because finite-size effects and noise remain in the MD data, we estimate $S^0_{\alpha\beta}=S_{\alpha\beta}(\mathbf{k}\rightarrow 0)$ by fitting the small-$\mathbf{k}$ data to the Ornstein--Zernike form and extrapolating to $\mathbf{k}=0$ (S0 method, \citet{Cheng_2022}).
We report the excess curvature $\tfrac{1}{k_{\mathrm B}T}\,\partial^2 g_{\mathrm{ex}}/\partial x^2$, where negative values signal a demixing tendency rather than a fully converged spinodal. As shown in the bottom panel of Fig.~\ref{fig:figure2}(a), this excess curvature turns negative near the FeB$_4$ composition at 1500~K, crossing zero around $x = 0.7$ and deepening toward the B-rich end. The structure-factor-derived demixing tendency thus corroborates the concavity of the fitted free energy, and because the finite cell truncates the long-wavelength fluctuations that would diverge at a singular Hessian, its magnitude likely underestimates the instability, consistent with the larger $S_{cc}(0)$ obtained in the 4320-atom cell.
Because the extrapolation of $S_{\mathrm{cc}}(0)$ is limited by the finite simulation-cell size, it carries a relatively large fitting uncertainty (Fig.~S5). 
We therefore repeated the FeB$_4$-composition simulation with 4320 atoms. The larger cell gives $S_{cc}(0)$ of the same order as the small cell at each pressure and preserves the pressure ordering, with the 10~GPa value remaining several times smaller than the 0~GPa value (Fig.~S5). 
Taken together, the free-energy curvature and the low-$k$ structure factor consistently indicate that pressure strongly suppresses the demixing tendency near the FeB$_4$ composition, while the precise location of the stability boundary at 10~GPa and 1800~K remains within our current uncertainty.

\noindent\textbf{Coordination numbers of B-rich phases.}
To identify the microscopic origin of this concavity, we turn to the local structure of the B-rich melt. 
B-rich borides commonly contain icosahedral and pentagonal-pyramidal motifs, in which boron is typically six-coordinated (CN=6) \cite{Albert_2009}; the CN=6 fraction therefore provides a structural proxy for low-energy B-rich local order. 
Motivated by this local geometry, we analyzed the coordination of B atoms in MD snapshots of Fe$_{0.1}$B$_{0.9}$, focusing on the B(CN=6) fraction and its correlation with the average potential energy (Fig.~\ref{fig:figure3}). At 1800~K, the CN=6 fraction at 10~GPa grows steadily with B content, whereas at 0~GPa it rises sharply above 85\% B, exceeding the 10~GPa value by about 10\% (Fig.~\ref{fig:figure3}(a)). This enhanced six-coordination at low pressure reflects a greater tendency to form icosahedral and pentagonal-pyramidal motifs (Fig.~\ref{fig:figure3}(b)), and its onset coincides with the turning point where the average potential energy begins to decrease with composition (dashed line, Fig.~\ref{fig:figure3}(a)). Under pressure, the melt favors denser packing and forms fewer such motifs. This shift is thermodynamically rationalized by the $PV$ term (Fig.~\ref{fig:figure3}(c)): comparing the average energy with and without the $PV$ correction at 1500~K and 10~GPa shows that the convex $PV$ contribution offsets the concavity of the potential-energy landscape. 
The reduced CN=6 signature under pressure, together with the convex $PV$ contribution, is consistent with a strongly reduced demixing tendency at the FeB$_4$ composition.

\begin{figure}[htbp]
    \centering
    \includegraphics[width=\columnwidth]{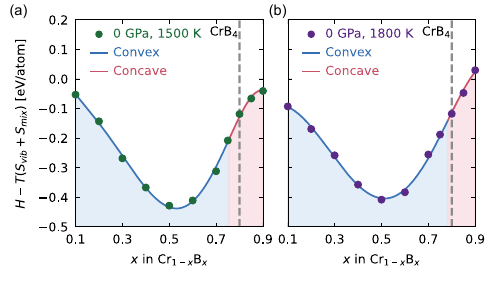}
    \caption{
    Free energy $H-T(S_\mathrm{vib}+S_\mathrm{mix})$ of the Cr--B system for (a) 1500 K and (b) 1800 K at 0 GPa, with regions colored by the sign of its curvature from a polynomial fit, classifying each composition as convex (in blue) or concave (in red).
    }
    \label{fig:figure4}
\end{figure}

\noindent\textbf{Cr--B interfacial melts.} CrB$_4$ provides a chemically distinct comparison because it is isostructural with FeB$_4$ but can be synthesized at ambient pressure~\cite{Kolmogorov_2010,CrB4_hardness}. Nevertheless, bulk CrB$_4$ remains difficult to obtain experimentally. For example, direct reaction of Cr and B with a Cr:B ratio of 1:4 remains incomplete even after annealing at 1523~K for up to 192~h, producing CrB$_4$ together with CrB$_2$ and residual amorphous boron~\cite{PhysRevB.85.144116}. Mineralizers or pressure are therefore commonly employed to facilitate its synthesis~\cite{CrB4_I,CrB4_pressure}. 
Applying the same melt free-energy analysis used for Fe--B, we find that the Cr--B and Fe--B free-energy curves show similar overall behavior (see Fig.~\ref{fig:figure4}), but the instability around the MB$_4$ composition is substantially weaker for CrB$_4$. At 1500 and 1800~K, the nearest stable Cr$_{1-x}$B$_x$ melt compositions deviate from CrB$_4$ by only $\Delta x=0.04$ and 0.01, respectively, compared with $\Delta x=0.18$ and 0.07 for FeB$_4$. Thus, although the CrB$_4$-composition melt is also weakly unstable at ambient pressure, the thermodynamic constraint on accessing the target composition is much smaller than for FeB$_4$. This difference is consistent with the experimental contrast: CrB$_4$ can be synthesized at ambient pressure after prolonged annealing, whereas FeB$_4$ generally requires high pressure. 
When pressure suppresses the melt instability, as our Fe--B simulations show and as is expected for Cr--B, the synthesis of both compounds is reported to complete within only tens of minutes~\cite{Gou_2013,CrB4_pressure}. The Cr--B system therefore provides an independent contrasting case linking the degree of melt instability to the experimentally observed synthesis difficulty.

\noindent\textbf{Discussion.}
Recent efforts to predict solid-state synthesis have moved beyond the thermodynamics of the final crystal toward the physics of the reactive interface \cite{shoemaker_situ_2014, McDermott_2021}, where a transient, liquid-like melt mediates product formation. \citet{Karan_2026} recently showed that correlated ionic transport through such a disordered interfacial layer governs the kinetic selectivity among competing products in the Ba--Ti--O system. Our framework addresses the complementary thermodynamic question for the same liquid-like region: whether a homogeneous melt at the target composition is locally stable, since a spinodally unstable melt cannot act as the parent phase from which the target nucleates. A convex hull for the crystalline product alone is therefore insufficient; the free-energy landscape of the interfacial melt should also be convex. Fe--B is a concrete illustration of how this single criterion accounts for a synthesis puzzle that has persisted since FeB$_4$ was first predicted.

The melt model used here to predict solid-state synthesizability is purely thermodynamic, whereas product nucleation is inherently kinetic. The two are linked as follows. When the model predicts no stable molten parent phase at the target composition, that composition cannot be accessed directly, and crystallization at the target composition then requires maintaining that composition against the demixing driving force. 
We do not invoke a barrier to demixing as spinodal decomposition is itself barrierless. 
What opposes crystallization is the free-energy penalty of holding the melt at the target composition, whose magnitude is set by $|G''|$. 
When that driving force is weak, the penalty is small, so the target phase can still form but is expected to evolve slowly, with a fine and texture-rich morphology as shown in the experiments (see Fig.~S38) \cite{Wang_2015_FeB}.

\begin{figure}[htbp]
    \centering
    \includegraphics[width=\columnwidth]{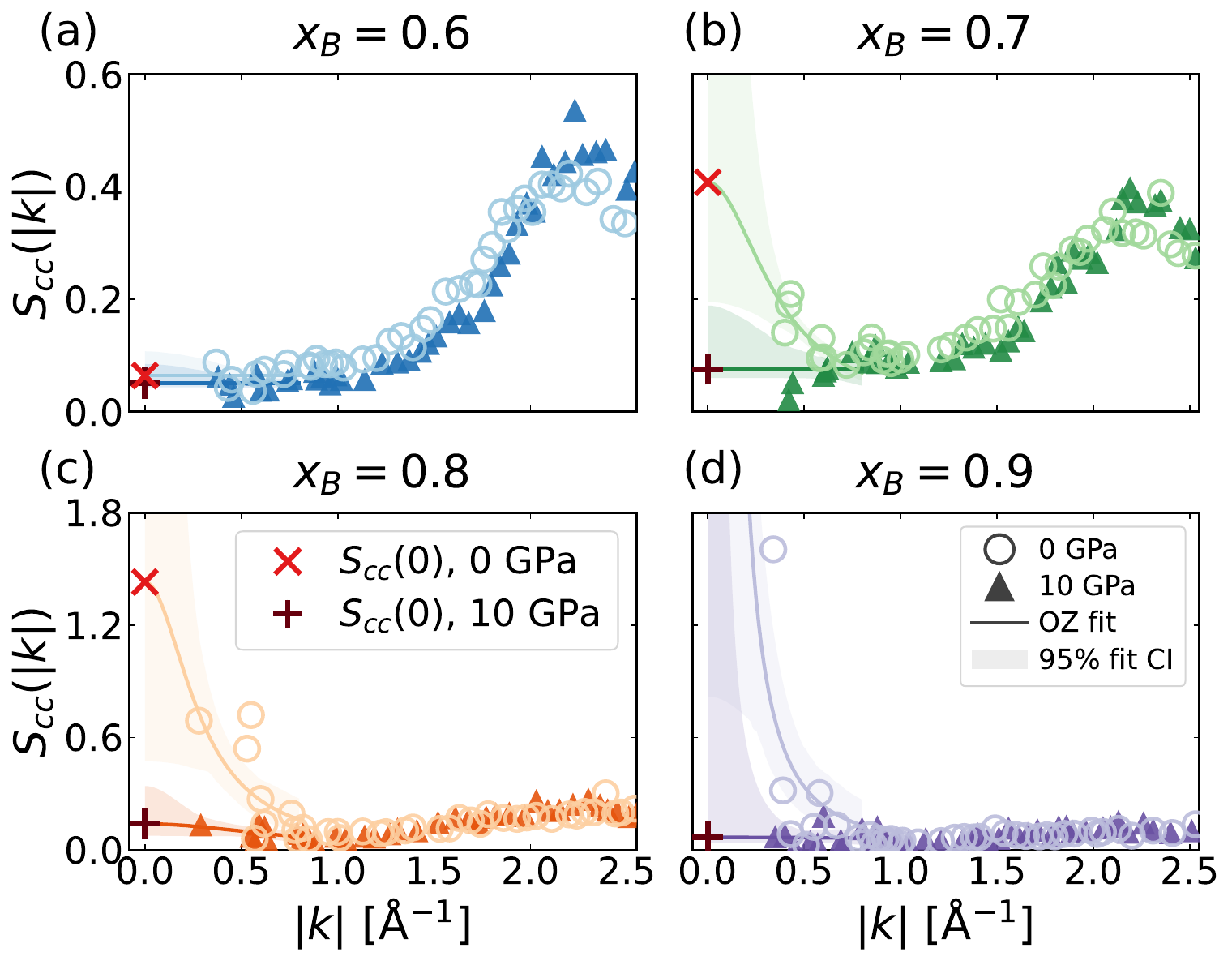}
    \caption{
    Concentration--concentration structure factor $S_{\mathrm{cc}}(k)$ at 1500~K for (a) $x_{\mathrm{B}}=0.6$, (b) $x_{\mathrm{B}}=0.7$, (c) $x_{\mathrm{B}}=0.8$, and (d) $x_{\mathrm{B}}=0.9$. The zero-wavevector limit satisfies $S_{\mathrm{cc}}(0)=x_{\mathrm{Fe}}x_{\mathrm{B}}/\Gamma$. Solid curves show weighted Ornstein--Zernike fits, and shaded regions indicate the corresponding 95\% confidence intervals. At 0~GPa, the low-$k$ enhancement becomes progressively stronger from $x_{\mathrm{B}}=0.7$ to 0.9, indicating increasing long-wavelength concentration fluctuations toward spinodal instability, consistent with the concave region of the energy fit in Fig.~\ref{fig:figure2}(a). This enhancement is strongly suppressed at 10~GPa.
    }
    \label{fig:figure5}
\end{figure}

Looking ahead, \textit{in silico} prediction must extend from crystal structures to synthesizability \cite{cheetham_chemical_2022, choi_synthesis-aware_2025, park_closing_2026}. By identifying the missing free-energy convexity in the interfacial melt as a possible origin of unsynthesizability, the framework offers a physically transparent basis for interpreting stable-but-unrealized compounds and for prioritizing experiments. Beyond fitting the energy landscape in composition space, we demonstrate that 
this instability can be diagnosed from the low-$k$ behavior of the concentration--concentration structure factor $S_{\text{cc}}(k)$. Fig.~\ref{fig:figure5} shows $S_{\text{cc}}(k)$ at 1500 K for Fe$_{1-x}$B$_x$ versus $|k|$: for $x > 0.6$ it rises steeply toward $k \to 0$ at ambient pressure and is markedly suppressed under 10~GPa, in agreement with the concave region from the curvature fit in Fig.~\ref{fig:figure2}(a). 
Because this signature arises from the small-$k$ structure factor in large-scale MD simulations rather than from the polynomial free-energy fit, it provides a practical descriptor of melt spinodal instability prior to experimental screening. 
Because mapping the complete melt--annealing free-energy landscape across the Fe--B composition space requires simulations at many compositions, these calculations employ smaller cells than would ideally be required for converging the low-$k$ structure factor, leading to a relatively large uncertainty in the extrapolated $S_{\mathrm{cc}}(0)$. Nevertheless, additional simulations containing 4320 atoms at the FeB$_4$ composition yield extrapolated values that preserve the pressure ordering found in the smaller cells (Fig.~S5), supporting the robustness of the structure-factor analysis.

Although the Fe--B case explains the pressure-dependent synthesis of FeB$_4$, achieving a predictive description of reactive interfaces remains an open challenge. The present demonstration is limited to a binary melt, whereas practical syntheses involve multicomponent systems with strong cation--anion correlations (e.g., complex oxides or polyanionic compounds) \cite{Chen_2024}. 
Generalizing the criterion to such multicomponent melts, which requires evaluating the full free-energy Hessian in the precursor compositions with \textit{ab initio} accuracy \cite{de_fontaine_analysis_1972}, will be essential for modeling realistic synthesis environments.

In summary, we propose and test a thermodynamic indicator criterion for interfacial-melt-mediated solid-state synthesis: beyond lying on the formation-energy convex hull, a target phase requires its interfacial melt to remain locally stable against spinodal decomposition. For Fe--B, this accounts for why the thermodynamically stable FeB$_4$ bulk has been reported under high pressure but not in low-pressure synthesis attempts.
At ambient pressure, the B-rich melt near the FeB$_4$ composition ($x_\text{B}=0.8$) develops a concave free-energy landscape and a strongly enhanced low-$k$ concentration--concentration structure factor, an instability correlated with low-energy icosahedral and pentagonal-pyramidal boron motifs. In contrast to FeB$_4$, the metastable Fe$_3$B and Fe$_{23}$B$_6$ phases lie higher above the convex hull yet remain experimentally synthesizable, because their corresponding B-deficient melts are stable, demonstrating that crystalline thermodynamic stability alone does not determine solid-state synthesizability. Applied pressure introduces a convex $PV$ contribution that suppresses these motifs and strongly reduces the demixing tendency, consistent with the experimental synthesis boundary. CrB$_4$, which is isostructural with FeB$_4$ but can be synthesized at ambient pressure, provides an independent comparison showing that the melt instability around the MB$_4$ composition is substantially weaker for CrB$_4$ than for FeB$_4$, consistent with the experimentally observed difference in synthesis difficulty and further supporting the relevance of melt stability to solid-state synthesizability.
This criterion provides a non-empirical complement to convex-hull stability. When integrated with kinetic models of interfacial transport, it can establish a predictive, experimentally aligned thermodynamic framework for screening solid-state synthesizability in AI-driven materials discovery.

\section*{Acknowledgments} This work was supported in part by the AI2050 program at Schmidt Sciences (Grant G-25-69776), the National Supercomputing Center of Singapore (NSCC), and the NUS High Performance Computing facility (CFP04-CF-033). We thank Shyue Ping Ong for valuable discussions.

\section*{Methods}

\noindent \textbf{MLIP fine-tuning.}~ Production simulations used a Fe--B fine-tuned MACE machine-learning interatomic potential (MLIP), an E(3)-equivariant message-passing model initialized from the MATPES r$^2$SCAN head of MACE-MH-1 \cite{batatia_cross_2025}.
The MACE-MH-1 weights were fine-tuned on 7582 Fe--B configurations drawn from MD trajectories over Fe$_{1-x}$B$_x$ ($0.1 \leq x \leq 0.9$), 1000, 1500, and 2000~K, and 0 and 10~GPa, then split 90:10 into training and held-out test sets.
Reference single-point energies and forces were computed with VASP using the projector augmented-wave method \cite{PAW_1994, Kresse_1996, Kresse_1999}, the r$^2$SCAN meta-GGA functional \cite{furness_2020}, and \texttt{MP24RelaxSet}. On the held-out test set, the potential reached mean absolute errors of 3.4~meV/atom in energy and 0.10~eV/\AA{} in force. Repeating the melt–quench workflow with the independent, pre-trained PBE-level OMAT head of MACE-MH-1 recovers the same curvature sign and pressure trend, indicating that these features are not specific to the fine-tuned potential (see SI). For the simulations of the Cr–B system, we adopted the MATPES r$^2$SCAN head from the pretrained MACE-MH-1 model to carry out melt-quench molecular dynamics. To probe interfacial melting in the Fe--B system, we constructed a two-phase supercell (288 Fe and 1152 B atoms) by joining the (010) surface of FeB with the (100) surface of $\alpha$-boron, and, after relaxation with the fine-tuned MACE-MH-1 model, performed NPT molecular dynamics at 1500~K and 0~GPa. 

\noindent \textbf{Melt--quench MD simulations.}~ Fe$_{1-x}$B$_x$ melts ($0.1 \leq x \leq 0.9$) were sampled at 0 and 10~GPa with a LAMMPS melt--quench MD protocol \cite{Aidan_2022}. We used AIRSS \cite{pickard_2011} to generate 20-atom homogeneous candidate structures with the MACE potential; for each composition, the lowest-energy candidate was expanded to a supercell of up to 540 atoms.
Each supercell was melted at 2600~K for 150~ps in the NPT ensemble, then cooled through 2300, 2000, 1800, 1500, and 1200~K, holding for 150~ps at each stage with a 1~fs time step and a Nos\'e--Hoover thermostat and barostat. Each stage started from the equilibrated configuration of the preceding stage, and structural and energetic analyses were performed on the second half of each trajectory. Full supercell sizes and schedules are provided in SI.

\noindent \textbf{Free-energy curvature from melt energetics.}~Melt stability was first assessed from the composition curvature of the per-atom free-energy profile $G(x)=H-T(S_\mathrm{vib}+S_\mathrm{mix})$.
To isolate mixing, each fitted profile was referenced to elemental simulations $E_{\mathrm{norm}}(\mathrm{Fe}_{1-x}\mathrm{B}_x) = E(\mathrm{Fe}_{1-x}\mathrm{B}_x) - (1-x)E(\mathrm{Fe}) - xE(\mathrm{B})$,
where $E$ denotes the per-atom quantity being normalized (potential energy, enthalpy, or free energy), and $E(\mathrm{Fe})$ and $E(\mathrm{B})$ were obtained from MD simulations initialized with elemental bcc-Fe and $\alpha$-B crystals at the same pressure and temperature.
Here $H$ is the trajectory-averaged enthalpy, $S_\mathrm{vib}$ was obtained from the vibrational density of states calculated from the velocity autocorrelation function \cite{lin_two-phase_2003}, and the configurational term was approximated by its ideal-solution upper bound
\begin{equation}
S_\mathrm{mix} = -k_{\mathrm B}(x_\text{Fe} \ln x_\text{Fe} + x_\text{B} \ln x_\text{B}).
\label{eq:mix}
\end{equation}
Because the ideal value overestimates the true configurational entropy of a clustering melt, it over-stabilizes the homogeneous state, so any residual concavity is a conservative indicator of instability. The curvature $G''(x)$ was obtained by differentiating a polynomial fit to the normalized free-energy profile.

\noindent \textbf{Structure factor and demixing.}~ As a structure-based estimator, we evaluated the demixing tendency from concentration fluctuations in the melt. For species $\alpha$ and $\beta$, the partial structure factors $S_{\alpha\beta}(k)$ were computed directly from the trajectories and combined into the Bhatia--Thornton concentration--concentration structure factor \cite{BhatiaThornton_1970},
\begin{equation}
S_{cc}(k)= x_\alpha x_\beta\left[x_\beta S_{\alpha\alpha}(k)+x_\alpha S_{\beta\beta}(k)-2\sqrt{x_\alpha x_\beta}\,S_{\alpha\beta}(k)\right],
\label{eq:scc}
\end{equation}
which is finite and positive for a homogeneous melt and diverges as $k\to0$ on approach to a spinodal. The composition curvature is set by the zero-wavevector limit
\begin{equation}
\Gamma=\frac{x_\alpha x_\beta}{S_{cc}(k\rightarrow0)},\qquad
\frac{\partial^2 g}{\partial x^2}\bigg|_{T,P}=\frac{k_{\mathrm B}T}{S_{cc}(k\rightarrow0)}.
\label{eq:gamma}
\end{equation}
The thermodynamic factor $\Gamma$ falls from unity for ideal mixing toward $0^+$ as the spinodal is approached.
Because a finite periodic box cannot resolve $k=0$ directly, $S_{cc}(0)$ was obtained by extrapolating the small-$k$ spectrum with an Ornstein--Zernike form \cite{barrat_basic_2003},
\begin{equation}
S_{cc}(k)=\frac{a}{1+bk^2},\qquad a=S_{cc}(0).
\label{eq:oz}
\end{equation}
Full derivations, the finite-size analysis, and the Kirkwood--Buff integral cross-check are provided in SI. 

\section*{Data availability}
Supporting data and source data for the figures are available at \url{https://github.com/AM3GroupHub/interfacial_melt}. 

\section*{Code availability}
The code base is open-sourced at \url{https://github.com/AM3GroupHub/interfacial_melt}. 

\section*{Author contributions}
P.Z. conceived the idea and designed the research. Z.Z. and M.C. performed the simulations and research. Z.Z., M.C. and P.Z. prepared the manuscript with help from Q.L. All authors contributed to discussions.

\section*{Competing interests}
The authors declare no competing interests.

\bibliography{references}
\end{document}